\documentclass{PoS}

\title{NLO QCD corrections to $Wb{\bar b}$ and $Zb{\bar b}$ production}

\ShortTitle{NLO QCD corrections to $Wb{\bar b}$ and $Zb{\bar b}$ production}

\author{Fernando Febres Cordero\\
        University of California, Los Angeles\\
        E-mail: \email{ffebres@physics.ucla.edu}}

\author{\speaker{Laura Reina}\\
        Florida State University\\
        E-mail: \email{reina@hep.fsu.edu}}

\author{Doreen Wackeroth\\
        University at Buffalo - The State University of New York\\
        E-mail: \email{dow@ubpheno.physics.buffalo.edu}}

\abstract{We present NLO QCD results for $Wb\bar{b}$ and $Zb\bar{b}$
production at the Tevatron including full bottom-quark mass effects.
We study the impact of QCD corrections on both total cross-section and
invariant mass distribution of the bottom-quark pair. Including NLO
QCD corrections greatly reduces the dependence of the tree-level
cross-section on the renormalization and factorization scales. We also
compare our calculation to a calculation that considers massless
bottom quarks and find that the bottom-quark mass effects amount to
about 8-10\% of the total NLO QCD cross-section and can impact the
shape of the bottom-quark pair invariant mass distribution, in
particular in the low invariant mass region.}

\FullConference{8th International Symposium on Radiative Corrections \\
                 October 1-5, 2007\\
                 Florence, Italy}

\begin{document}

\section{Introduction}
\label{sec:intro}

Precise theoretical predictions for the production of a weak gauge
boson ($W/Z$) in association with a $b\bar{b}$ pair are very important
both in the search for a light Standard Model (SM)-like Higgs boson
and in the search for single-top production at hadron colliders, in
particular at the Tevatron. Indeed, $W/Z+b\bar{b}$ represent the major
irreducible backgrounds to the associated production of a light Higgs
boson with a weak gauge boson ($WH$ and $ZH$ with $H\rightarrow
b\bar{b}$), which are the main search modes for a SM-like Higgs boson
at the Tevatron~\cite{Abazov:2007h,abulencia:2006ep}.  At the same
time, $Wb\bar{b}$ is an irreducible background for single-top
production, which is being measured for the first time at the Tevatron
as $p\bar{p}\rightarrow t\bar{b},\bar{t}b$ with $t(\bar{t})\rightarrow
b(\bar{b})W$~\cite{Abazov:2007ev,Acosta:2004bs}.

The signal cross sections for both $WH/ZH$ and single-top production
are known including higher order QCD and electroweak corrections. It
is therefore important to also control the background to a good level
of theoretical precision. In the present experimental analyses the
effects of Next-to-Leading (NLO) QCD corrections on the total
cross-section and the dijet invariant mass distribution of the
$Wb\bar{b}$ and $Zb\bar{b}$ background processes have been taken into
account by using the MCFM package~\cite{MCFM:2004}, which implements
the zero bottom-quark mass ($m_b=0$)
approximation~\cite{Ellis:1998fv,Campbell:2000bg,Campbell:2002tg}.

In this proceedings we present results for the NLO QCD cross-sections
and $b\bar{b}$-pair invariant mass distributions at the Tevatron,
including full bottom-quark mass effects~\cite{Febres
Cordero:2006sj}. We separately study the impact of NLO QCD corrections
and of non-zero bottom-quark mass effects. NLO QCD corrections
stabilize the theoretical prediction of total cross-sections and
distributions, reducing the dependence on the renormalization and
factorization scales. On the other hand, the presence of a non-zero
bottom-quark mass mainly affect the cross-section in the region where
the $b\bar{b}$-pair invariant mass is small, both at Leading Order
(LO) and at NLO in QCD, and amount to an overall 8-10\% difference
with respect to the zero bottom-quark mass approximation.

\section{NLO calculation}
\label{sec:nlo_calc}
 
The hadronic production of a $W$ boson with a $b\bar{b}$ pair occurs
at tree level in QCD via the $q\bar{q}^\prime\rightarrow Wb\bar{b}$
partonic process. On the other hand, the hadronic production of a $Z$
boson with a $b\bar{b}$ pair consists, at the tree level in QCD, of
two partonic channels, namely $q\bar{q}\rightarrow Zb\bar{b}$ and
$gg\rightarrow Zb\bar{b}$. The NLO QCD corrections to the tree level
partonic cross-section consists of both one-loop $O(\alpha_s)$
virtual corrections and $O(\alpha_s)$ real corrections
corresponding to the emission of one extra parton from the tree level
parton processes. The NLO hadronic cross-section is obtained by convoluting
the parton-level NLO cross-sections with NLO Parton Distribution
Functions (PDF).

The $O(\alpha_s)$ virtual corrections to the partonic
cross-section contain ultraviolet (UV) and infrared (IR) singularities.
The UV singularities are calculated using dimensional regularization
and cancelled by introducing a suitable set of counterterms (see
Ref.~\cite{Febres Cordero:2006sj} for details). IR singularities are
isolated using dimensional regularization and cancelled against the
analogous singularities arising in the $O(\alpha_s)$ real
corrections to the partonic cross-section.

The $O(\alpha_s)$ virtual corrections to the partonic cross-section
consist of one-loop self-energy, vertex, box and pentagon diagrams. We
apply techniques developed in the NLO QCD calculation of
$Ht\bar{t}$~\cite{Reina:2001bc,Dawson:2003zu} for the calculation of
scalar and tensor loop-integrals. In particular, we calculate the
tensor loop-integrals via Passarino-Veltman reduction
(PV)~\cite{Passarino:1989ey}. We encounter instabilities due to the
quasi-vanishing of the Gram determinant(s) of the process only in one
box diagram and in several pentagon diagrams. We are able to obtain
stable numerical results by combining, when necessary, sets of gauge
invariant diagrams.

The $O(\alpha_s)$ real corrections to the partonic cross-section
consist of the $q\bar{q}^\prime\rightarrow Wb\bar{b}+g$ and
$qg(\bar{q}g)\rightarrow Wb\bar{b}+q(\bar{q})$ subprocesses for
$Wb\bar{b}$ production and of the $q\bar{q}\rightarrow Zb\bar{b}+g$,
$gg\rightarrow Zb\bar{b}+g$, and $qg(\bar{q}g)\rightarrow
Zb\bar{b}+q(\bar{q})$ subprocesses for $Zb\bar{b}$ production.  We
have extracted both soft and collinear IR singularities by
implementing a Phase Space Slicing method with two
cutoffs~\cite{Harris:2001sx} in order to isolate the soft ($\delta_s$)
and collinear ($\delta_c$) singularities respectively.  Both the soft,
hard-collinear, and hard-non-collinear parts of the cross-section
depend on the cutoffs, but their sum is cutoff independent over a
large range of values of the cutoffs (see Ref.~\cite{Febres
Cordero:2006sj}). 

Both analytical and numerical results for the NLO hadronic
cross-section have been checked with two independent calculations
based on different programming languages and public/in-house packages.
The analytical reduction of the calculation has been obtained using
FORM~\cite{Vermaseren:2000nd} and
\emph{Maple} codes, while the numerical results have been obtained using
Fortran and C codes. The FF package~\cite{vanOldenborgh:1990yc} has
been used to check some of the IR-finite scalar and tensor
integrals. Finally, the hard non-collinear real corrections have been
double-checked using
Madgraph~\cite{Murayama:1992gi,Stelzer:1994ta,Maltoni:2002qb}.

\section{Numerical results}
\label{sec:num_res}

In these proceedings we present results for $Wb\bar{b}$ and
$Zb\bar{b}$ production at the Tevatron including NLO QCD corrections
and a non-zero bottom-quark mass fixed at $m_b\!=\!4.62$~GeV. Results
for $Wb\bar{b}$ have been published in Ref.~\cite{Febres
Cordero:2006sj}, while results for $Zb\bar{b}$ are currently being
cross-checked and should be considered as preliminary. Both $W$ and
$Z$ boson are considered on-shell and their masses are taken to be
$M_W=80.41$~GeV for the $Wb\bar{b}$ runs and $M_Z=91.1876$~GeV for the
$Zb\bar{b}$ runs, while, in each case, the other weak gauge boson mass
is calculated via the relation $M_W=\cos\theta_wM_Z$ with
$\sin^2\theta_w=0.223$.  The LO results use the 1-loop evolution of
$\alpha_s$ and the CTEQ6L set of PDF~\cite{Lai:1999wy}, while the NLO
results use the 2-loop evolution of $\alpha_s$ and the CTEQ6M set of
PDF, with $\alpha_s^{NLO}(M_Z)=0.118$.  The $W$ boson coupling to
quarks is proportional to the Cabibbo-Kobayashi-Maskawa (CKM) matrix
elements.  We take $V_{ud}=V_{cs}=0.975$ and $V_{us}=V_{cd}=0.222$,
while we neglect the contribution of the third generation, since it is
suppressed either by the initial state quark densities or by the
corresponding CKM matrix elements.
 
We implement the $k_T$ jet
algorithm~\cite{Catani:1992zp,Catani:1993hr,Ellis:1993tq,Kilgore:1996sq}
with a pseudo-cone size $R=0.7$ and we recombine the parton momenta
within a jet using the so called covariant
$E$-scheme~\cite{Catani:1993hr}. We checked that our implementation of
the $k_T$ jet algorithm coincides with the one in MCFM.  We require
all events to have a $b\bar{b}$ jet pair in the final state, with a
transverse momentum larger than $15$~GeV ($p_T^{b,\bar{b}}>15$~GeV)
and a pseudorapidity that satisfies $|\eta^{b,\bar{b}}|<2$. We impose
the same $p_T$ and $|\eta|$ cuts also on the extra jet that may arise
due to hard non-collinear real emission of a parton, i.e. in the
processes $W/Zb\bar{b}+g$ or $W/Zb\bar{b}+q(\bar{q})$. This hard
non-collinear extra parton is treated either \emph{inclusively} or
\emph{exclusively}, following the definition of \emph{inclusive} and
\emph{exclusive} as implemented in the MCFM code~\cite{MCFM:2004}.  In
the \emph{inclusive} case we include both two- and three-jet events,
while in the \emph{exclusive} case we require exactly two jets in the
event. Two-jet events consist of a bottom-quark jet pair that may also
include a final-state light parton (gluon or quark) due to the applied
recombination procedure. Results in the massless bottom-quark
approximation have been obtained using the MCFM code~\cite{MCFM:2004}.

\begin{figure}[htb]
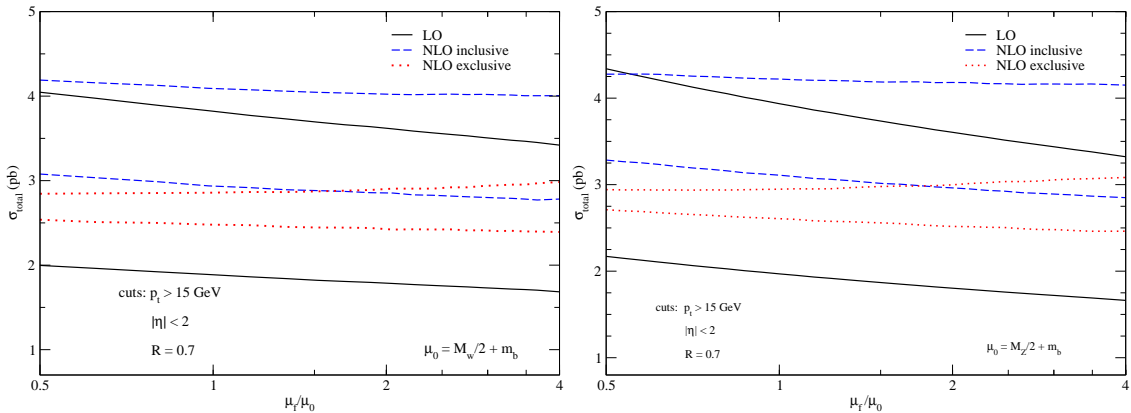

\begin{center}
\includegraphics*[scale=0.32]{bands_mur_muf_Wbb} 
\includegraphics*[scale=0.32]{bands_mur_muf_Zbb} 
\caption[]{Dependence of the LO (black solid band), NLO
\emph{inclusive} (blue dashed band), and NLO
\emph{exclusive} (red dotted band) total cross-sections on the
renormalization/factorization scales, including full bottom-quark mass
effects. The l.h.s. plot is for $p\bar{p}\rightarrow Wb\bar{b}$ and
the r.h.s. plot for $p\bar{p}\rightarrow Zb\bar{b}$ .  The bands are
obtained by varying both $\mu_R$ and $\mu_F$ independently between
$\mu_0/2$ and $4\mu_0$ (with $\mu_0=m_b+M_V/2$ for $V=W,Z$ in the
$p\bar{p}\rightarrow Wb\bar{b}$ and $p\bar{p}\rightarrow Zb\bar{b}$
cases respectively).}
\label{fig:mu_dependence_band}
\end{center}
\end{figure}
\begin{figure}[htb]
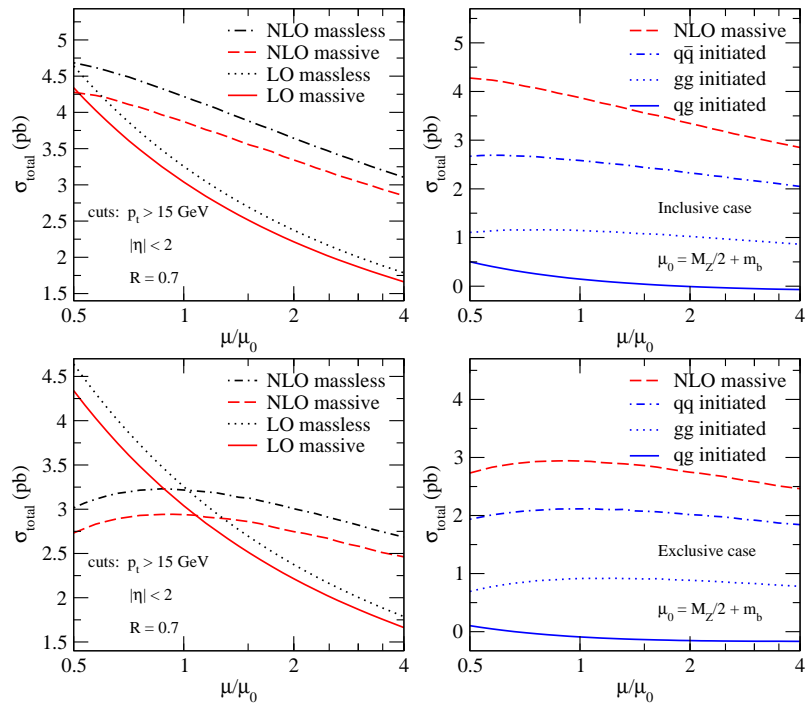

\begin{center}
\includegraphics*[scale=0.45]{mu_dependenceInc_Zbb} 
\includegraphics*[scale=0.45]{mu_dependenceExc_Zbb} 
\caption[]{Dependence of the LO and NLO \emph{inclusive} (upper plots) and
\emph{exclusive} (lower plots) total cross-section for $p\bar{p}\rightarrow Zb\bar{b}$
on the renormalization/factorization scale, when
$\mu_R\!=\!\mu_F$. The l.h.s. plots compare both LO and NLO total
cross-sections for the case in which the bottom quark is treated as
massless (MCFM) or massive (our calculation).  The r.h.s.  plots show
separately, for the massive case only, the scale dependence of the
$q\bar{q}$, $gg$ and $qg+\bar{q}g$ contributions, as well as their
sum.}
\label{fig:mu_dependence_Zbb}
\end{center}
\end{figure}

In Fig.~\ref{fig:mu_dependence_band} we
illustrate the renormalization and factorization scale dependence of
the LO and NLO total cross-sections, both in the \emph{inclusive} and
\emph{exclusive} case.  The bands are obtained by varying both $\mu_R$ and $\mu_F$
independently between $\mu_0/2$ and $4\mu_0$ (with $\mu_0=m_b+M_V/2$
for $V=W,Z$ in the $p\bar{p}\rightarrow Wb\bar{b}$ and
$p\bar{p}\rightarrow Zb\bar{b}$ cases respectively), including full
bottom-quark mass effects. We notice that the NLO cross-sections have
a reduced scale dependence over most of the range of scales shown, and
the \emph{exclusive} NLO cross-section is more stable than the
\emph{inclusive} one especially at low scales. While the
LO cross-section still has a 40\% uncertainty due to scale dependence,
this uncertainty is reduced at NLO to about 20\% for the
\emph{inclusive} and to about 10\% for the \emph{exclusive}
cross-section respectively. This is consistent with the fact that the
\emph{inclusive} NLO cross-section integrates over the entire phase
space of the $qg(\bar{q}g)\rightarrow b\bar{b}W/Z + q(\bar{q})$
channels that are evaluated with NLO $\alpha_s$ and NLO PDF, but are
actually tree-level processes and retain therefore a strong scale
dependence. In the \emph{exclusive} case only the $2\rightarrow 3$
collinear kinematic of these processes is retained, since 3-jets
events are discarded, and this makes the overall renormalization and
factorization scale dependence milder. This is better illustrated in
Fig.~\ref{fig:mu_dependence_Zbb} for the case of $p\bar{p}\rightarrow
Zb\bar{b}$, where the r.h.s. plots show the scale dependence of the
cross-sections due to the single partonic channels ($q\bar{q}$, $gg$
and $qg+\bar{q}g$). The strong residual scale dependence of the
\emph{inclusive} NLO cross-section is clearly driven by the
$qg+\bar{q}g$ channel. Similar results are obtained for
$p\bar{p}\rightarrow Wb\bar{b}$, as reported in Ref.~\cite{Febres
Cordero:2006sj}.

\begin{figure}[htb]
\begin{center}
\includegraphics*[scale=0.4,angle=-90]{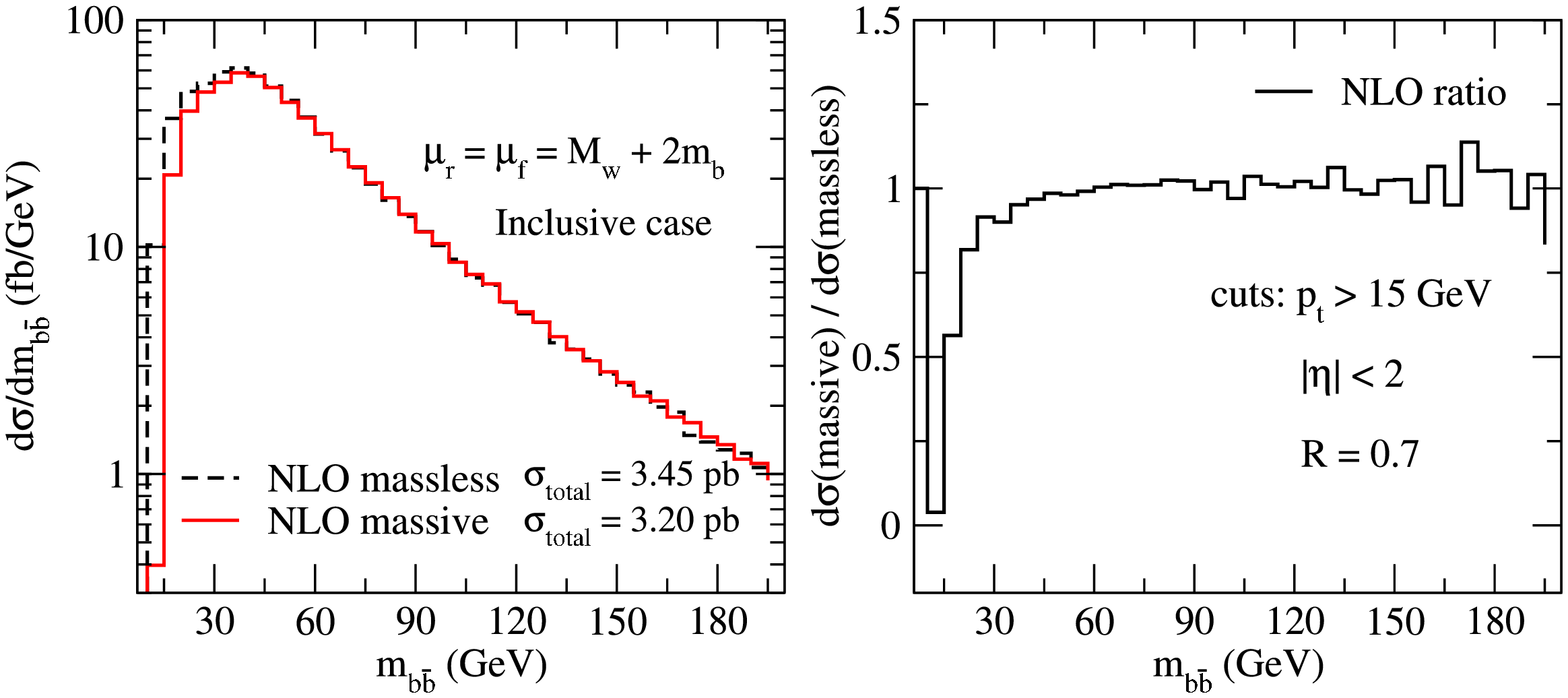} 
\includegraphics*[scale=0.4,angle=-90]{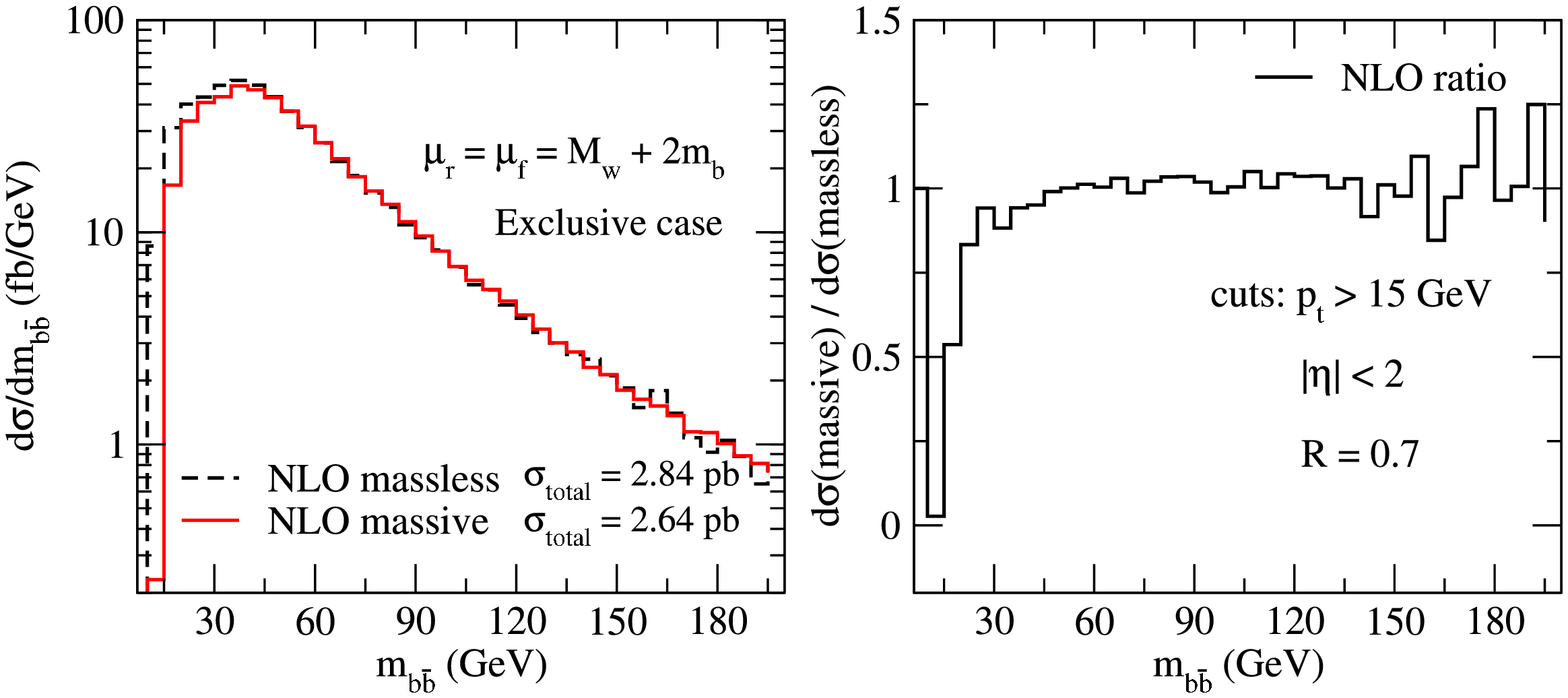} 
\caption[]{The \emph{inclusive} (upper plots) and \emph{exclusive} (lower plots) 
distributions, $d\sigma/dm_{b\bar{b}}$, for $p\bar{p}\rightarrow
Wb\bar{b}$ derived from our calculation (with $m_b\ne 0$) and from
MCFM (with $m_b=0$).  The right hand side plot shows the ratio of the
two distributions, $d\sigma(m_b\neq 0)/d\sigma(m_b=0)$. From
Ref.~\cite{Febres Cordero:2006sj}.}
\label{fig:NLO_massless_vs_massive}
\end{center}
\end{figure}

A first illustration of the impact of keeping a non-zero bottom-quark
mass in the calculation of the NLO QCD cross-section is given in the
l.h.s. plots of Fig.~\ref{fig:mu_dependence_Zbb}, where LO and NLO
total cross-sections for $p\bar{p}\rightarrow Zb\bar{b}$ are given,
both for $m_b=0$ and $m_b=4.62$~GeV, as functions of the
renormalization and factorization scales (identified for the purpose
of this plot). Neglecting bottom-quark mass effects overestimate the
NLO cross-section by about 8-10\%, depending on the choice of the
scale. In Fig.~\ref{fig:NLO_massless_vs_massive} we analyze the impact
of a non-zero bottom-quark mass on the $b\bar{b}$-pair invariant-mass
($m_{b\bar{b}}$) distribution. We give results for both
\emph{inclusive} and \emph{exclusive} distribution in the
$p\bar{p}\rightarrow Wb\bar{b}$ case. In both cases
distributions most of the impact is in the low $m_{b\bar{b}}$ region,
which can be important in a variety of different analyses.

\end{document}